\RequirePackage{fix-cm}
\documentclass[review]{elsarticle}
\usepackage[table]{xcolor}
\usepackage{appendix}
\usepackage{amsmath}
\usepackage{array}
\usepackage{longtable}
\usepackage{hyperref}
\usepackage[utf8]{inputenc}
\usepackage{graphicx}
\usepackage{amsfonts}
\usepackage{rotating}
\usepackage[singlespacing]{setspace}
\usepackage{fancyhdr}
\usepackage{tikz}
\usetikzlibrary{shapes}
\usetikzlibrary{positioning}
\usepackage{natbib}

\usepackage{textcomp}
\usepackage{booktabs}
\usepackage[final]{changes}

\usepackage{bm} 
\usepackage{amsfonts}
\usepackage{amssymb}
\usepackage{mathtools}
\usepackage{stmaryrd}
\usepackage{mathrsfs}
\newcommand{\bs}{\boldsymbol}
\usepackage{standalone}

\definecolor{Gray}{gray}{0.9}

\begin{document}
\begin{frontmatter}
\title{Observing water level extremes in the Mekong River Basin: The benefit of long-repeat orbit missions in a multi-mission satellite altimetry approach
}



\author[adress]{Eva Boergens \corref{mycorrespondingauthor}}
\cortext[mycorrespondingauthor]{Corresponding author}
\ead{eva.boergens@tum.de}
\author[adress]{Denise Dettmering}

\author[adress]{Florian Seitz}

\address[adress]{Deutsches Geodätisches Forschungsinstitut, Technische Universität München\\Arcisstr. 21\\80333 München}



\begin{abstract}
Single-mission altimetric water level observations of rivers are spatially and temporally limited, and thus they are often unable to quantify the full extent of extreme flood events. Moreover, only missions with a short-repeat orbit, such as Envisat, Jason-2, or SARAL, could provide meaningful time series of water level variations directly. However, long or non-repeat orbit missions such as CryoSat-2 have a very dense spatial resolution under the trade-off of a repeat time insufficient for time series extraction. Combining data from multiple altimeter missions into a multi-mission product allows for increasing the spatial and temporal resolution of the data. In this study, we combined water level data from CryoSat-2 with various observations from other altimeter missions  in the Mekong River Basin between 2008 and 2016 into one multi-mission water level time series using the approach of universal kriging. In contrast to former multi-mission altimetry methods, this approach allows for the incorporation of CryoSat-2 data as well as data from other long or non-repeat orbit missions, such as Envisat-EM or SARAL-DP. Additionally, for the first time, data from tributaries are incorporated. The multi-mission time series including CryoSat-2 data adequately reflects the general inter-annual flood behaviour and the extreme floodings in 2008 and 2011. It performs better than single-mission time series or multi-mission time series based only on short-repeat orbit data. The Probability of Detection of the floodings with the  multi-mission altimetry was around 80\% while Envisat and Jason-2 single-mission altimetry could only detect around 40\% of the floodings correctly. However, small flash floods still remain undetectable. \par
\end{abstract}

\end{frontmatter}
\section{Introduction}

The Mekong River Basin in South-East Asia is well known for its high and stable annual floods \citep{Adamson2009}. Floods are essential to the livelihoods of riparians, providing irrigation to the paddy fields, other agricultural activities and seasonal fisheries \citep{MRC2010}. Lower than average river levels in the flood season can lead to water shortages in the basin during the following dry season between December and May \citep{MRC2005}. On the other side, severe floodings can destroy infrastructure and agriculture. Despite the need to monitor river stages to forecast floods and identify long-term changes, the availability of global in situ gauge data has decreased in the past decades \citep{GRDC2013}.

An increasing number of studies have used satellite altimetry to measure river water levels independently from in situ observations with satellite altimetry \citep[e.g.][]{Birkinshaw2010, daSilva2010, Schwatke2015}, so that small rivers ($<$200\,m wide) can now be accurately observed \citep[e.g.][]{Maillard2015,Boergens2016,Biancamaria2016,Huang2018}. The launch of CryoSat-2 in 2010 increased the accuracy of water level observations of small rivers even further due to its Synthetic Aperture Radar (SAR) altimeter which has a smaller along-track footprint, compared to pulse limited altimetry. This technology has enabled higher measurement accuracy for small rivers less than 200\,m in width \citep{Villadsen2015, Boergens2017b}.

Data from short-repeat missions such as Envisat, SARAL, Jason-2 and Jason-3 can be used to build water level time series at a given location (i.e. virtual station; VS). However, the spatial and temporal resolution of satellite altimetry datasets is limited by the orbit design of satellite altimetry missions. Since 2010, at least three altimeter missions have been available to simultaneously observe water levels. Each mission has specific limitations with respect to temporal and spatial resolution. Even with the 10-day temporal resolution of Jason-2 and Jason-3, many flooding events cannot be observed in a river basin. Many missions available since 2010 have long or non-repeat orbits (CryoSat-2, Envisat-EM, and SARAL-DP), so time series at VSs cannot be estimated. But these missions, \replaced{mostly}{most} CryoSat-2, provide very valuable information for observing the flood dynamics of rivers and lakes, especially in the data gap between Envisat and SARAL in 2011 and 2012. Thus, it is necessary to combine data from short, long and non-repeat orbit missions into a multi-mission altimetry dataset. Multi-mission combination is already operational for lakes and reservoirs \citep{Schwatke2015}, where multiple altimeter missions and tracks are combined under the assumption that all measurements observe the same equipotential water surface. 

Unlike for lakes and reservoirs, combining data along rivers requires knowledge of river topography, slope and flow velocity \citep{Villadsen2015}. So far three studies have been published addressing multi-mission altimetry along rivers, with only one including CryoSat-2. The multi-mission river altimetry study by \mbox{\citet{Tourian2016}} have investigated altimetry data densification in the main stream of the Po River in Italy using data from Jason-2, Envisat, SARAL and CryoSat-2. Data combination has been achieved with the flow velocity between observations based on river slope and width which were derived from in situ data and Landsat 7 images. Thus, auxiliary information have been necessary for the combination of altimetry data.

In \mbox{\citet{Tourian2017}}, water levels from Jason-2, Envisat and SARAL have been combined along the Niger River and two major tributaries in West Africa. Following \mbox{\citet{Tourian2016}}, altimetric water levels have been densified to the location of in situ discharge stations at which a discharge time series have been estimated from the altimetry data.  A Kalman filter has been employed to assimilate the discharge time series of all in situ stations into a linear dynamic model with a stationary stochastic model. Due to the transformation to discharge, the topography along the river did not need to be further considered.

\mbox{\citet{Boergens2017}} have employed ordinary kriging to combine Envisat, SARAL and Jason-2 water level time series along the main stretch of the Mekong River. A prerequisite of ordinary kriging is a stochastic model mirroring the rivers flow. In the study two spatio-temporal covariance models have been presented: a stationary and non-stationary. However, the ordinary kriging approach did not allow for including of data from long or non-repeat orbit missions such as CryoSat-2 because it requires a reduction of the mean water level before the combination. With long or non-repeat orbit missions the mean water level is unknown and topography models have an insufficient accuracy for this reduction.

The present study aims at extending \added{and improving} the approach by \mbox{\citet{Boergens2017}} in order to include the valuable data of long- and non-repeat orbit missions, most notably CryoSat-2, and increases the study area also to the tributaries of the main river. In this context the three major research questions of this study are:
\begin{itemize}
 \item \textbf{What is the benefit  of altimetry missions on long and non-repeat orbits for river flood monitoring? }
 \item \textbf{How can measurements along tributaries be included? How do they influence the quality of water level interpolation? }
 \item \textbf{How well can multi-mission altimetry quantify inter-annual flood variations? Which are the most important factors influencing their accuracy and reliability? }
\end{itemize}

The previously applied multi-mission approach based on ordinary kriging is not adequate to answer these questions due to the reasons given above. For the problem at hand, we have to apply the generalised  approach of universal kriging \mbox{\citep{Delfiner1975,Myers1982}}. 
Universal kriging has been employed in hydrology before \replaced{interpolating}{to interpolate} ground water tables  \citep{Kumar2007,Gundogdu2007}  and precipitation levels \citep{Kastelec2002}. \citet{Brus2007} applied universal kriging to multiple environmental variables. 

Unlike ordinary kriging, universal kriging does not rely on the assumption of a constant data mean \citep{Cressie1993}. Thus a reduction of topography is not necessary---a pre-requisite for the incorporation of data from CryoSat-2, Envisat-EM, or SARAL DP where the mean water level or topography cannot be derived from an altimetry time series. This study inherits the non-stationary covariance models from \citet{Boergens2017} but new model parameters are estimated based on the extended data set.  This covariance model allows to include water level observation of the tributaries as well. The influence of the additional information sets will be evaluated in the study.

\section{Study Area}
\label{sec:area}
\begin{figure}
\centering
 \includegraphics[width=\textwidth]{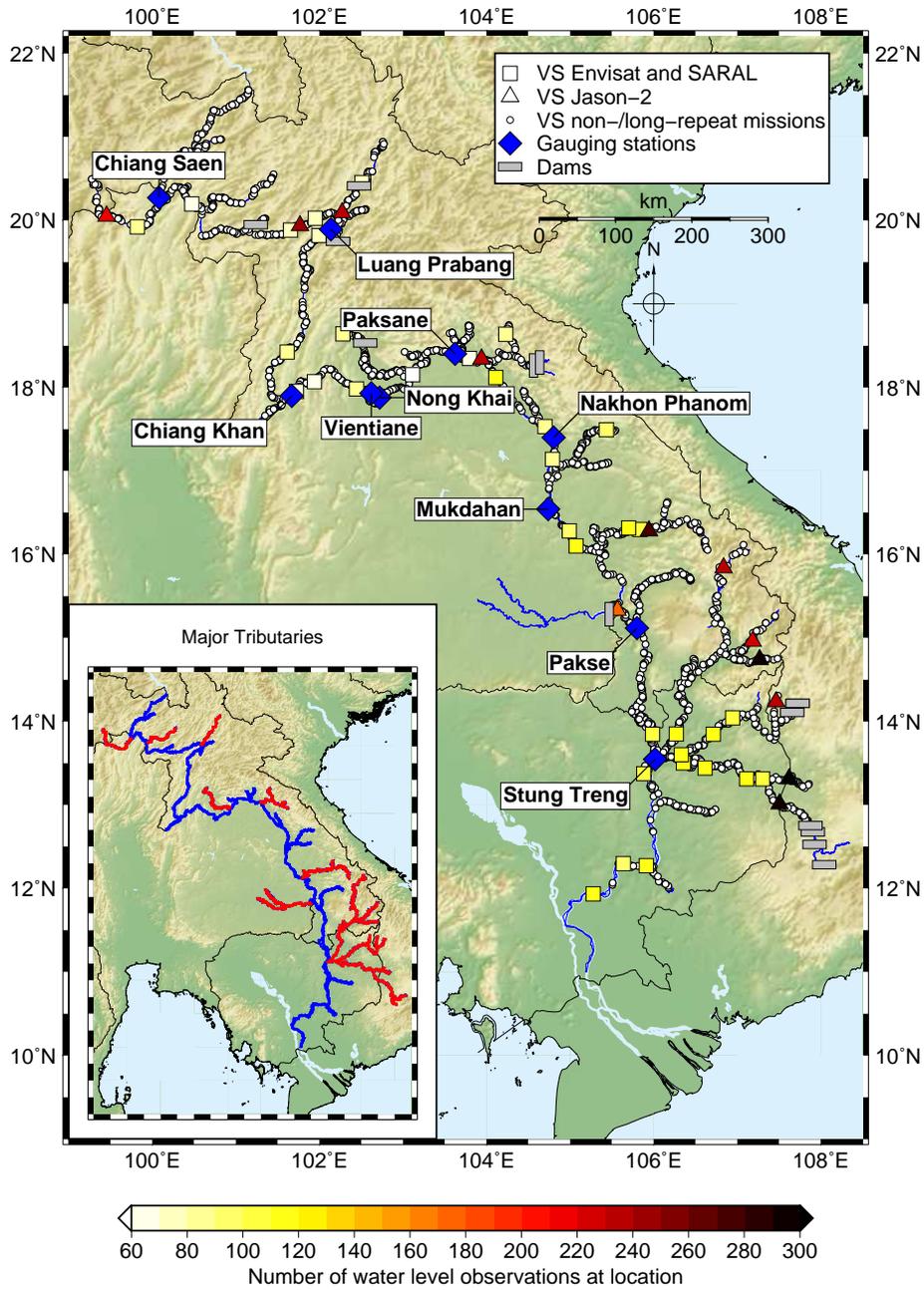}
 \caption{Map of the study area. The locations of altimetric water level observations are plotted, with colours indicating the number of observations at a location. For the long or non-repeat missions, the maximum number of observations \added{ at the same location} is six (CryoSat-2). Dams considered in this study are indicated. The inlay map shows major tributaries in red.}
 \label{fig:map}
\end{figure}

This work constitutes a case study of the Mekong River Basin in South-East Asia (\autoref{fig:map}). We investigated the river reach between the Chinese border and the confluence with the Tonle Sap River in Phnom Penh, including its tributaries. The river channels of the northern reach in China are too steep to use satellite altimetry \added{(see \citet{Huang2018} for details on problems with steep river gorges)}. Downstream of Phnom Penh, the Mekong widens into a delta with tidal influence.  Along some of the tributaries dams have been built. In such cases only the river reaches downstream of dams are considered in this study. 

The hydrology of the Mekong is dominated by two major compartments\citep{Adamson2009, MRC2005}. The first compartment is called the Yunnan compartment and is governed by snow melt on the Tibetan Plateau. The discharge of this compartment governs the flow of the river up to Vientiane, and constitutes up to 30\% of the average dry season flow of the Mekong Basin. The main flood of the Yunnan compartment occurs in August and September. South of Vientiane, the South-Eastern monsoon drives the hydrology of the Mekong and thus the monsoon compartment. The major left-bank tributaries in Laos, entering the main river between Vientiane and Stung Treng, are only governed by the monsoon and provide 50\% of the overall runoff of the Mekong River Basin. The monsoon lasts from mid-May to mid-October, defining the annual flood season during June and November, with its main peak in precipitation and water level in August. The water level variations of the main river can be as high as 10\,m \citep[e.g.][]{Boergens2016}.

\section{Data}
\label{sec:data}

\subsection{Altimetry Data}
\label{sec:alt_data}

The multi-mission altimetry approach used in this study relies on data from altimeter missions with a short-repeat cycle, and data from missions with either long repeat times or non-repeat/drifting orbits. Missions with short repeat times include Envisat (2002–2010) and its successor in the same orbit SARAL (2013–2016), with repeat times of 35 days. This leads to a VS being located approximately every 70\,km along the river. The altimeter missions of Jason-2 (2008–2016) and its successor Jason-3 (2016-present) have repeat times of only 10 days, leading to an even sparser spatial coverage. The long repeat or non-repeat missions of CryoSat-2 (2010–present), Envisat-EM (2010–2011) and SARAL-DP (2016–present) show very high spatial resolutions, but temporal resolution is limited to 369 days for CryoSat-2, while Envisat-EM and SARAL-DP have drifting orbits and therefore do not have repeat observations. CryoSat-2 observes the study area in both Synthetic Aperture Radar (SAR) and pulse limited Low Resolution (LRM) mode.

High-frequency data from missions with short-repeat times (Envisat, SARAL, Jason-2 and Jason-3) are processed in DAHITI (Database for Hydrological Time Series of Inland Waters) to create water level time series \citep{Schwatke2015}.  In DAHITI, the water level time series are derived from the altimeter observations in the vicinity of the river. In a first step an outlier removal based on the along-track standard deviation is applied; the resulting observations are then combined with a Kalman filter to a time series. The water level observations of the long or non-repeat orbit missions measuring in pulse limited mode, namely Envisat EM, SARAL DP and CryoSat-2 LRM, are derived at every crossing of an altimeter track with a river in the basin. To this end, if applicable a hooking effect correction is applied \citep{Boergens2016}, which is mostly for the crossings along small tributaries. In all other cases, the outlier detection of DAHITI is applied to all high-frequency data in the river's vicinity and the water level is estimated as the median of the residual altimeter observations. However, not every crossing delivers a usable water level observation.

Water levels from CryoSat-2 SAR data are computed based on a classification approach \citep{Boergens2017b}. In this approach, CryoSat-2 stack data \mbox{\citep{Wingham2006}} is used to derive classification features for all altimeter observations in the vicinity of 20\,km to any river in the Mekong Basin. With these features the data is classified with a k-means classifier \replaced{ into 20 clusters. Following, these clusters are then assigned to land and water observation classes. }{ in water and land observations}. The water observations are then used at each crossing of the altimeter track with a river to estimate the water level at this location. 

\added{The outlier removal for the long- and non-repeat orbit missions consists of two parts of which the first is only applicable for CryoSat-2 data. It is based on a comparison to water level observations of the same track but one year apart and on a comparison to water levels of other tracks in close spatial and temporal vicinity. The outlier detection as well as the whole data processing of CryoSat-2 SAR data is described in detail in  \citet{Boergens2017b}.}

\added{To assure consistency between all observations of the different missions, the same retracker, Improved Threshold Retracker \citep{Gommenginger2011}, was used to derive water level observations.}
For all missions, it is ensured that the same atmospheric and geophysical corrections (wet and dry troposphere, ionosphere, earth and pole tide, and geoid) and retrackers are used to create consistent data sets.  To combine altimeter missions it is necessary to correct for radial orbit offsets, which can cause height difference between water level observations from different missions \citep{Bosch2014}. All data upstream from dams are discarded (see \autoref{sec:area}).

\autoref{fig:data} shows the temporal availability of the different missions. The lower panel of this figure displays the number of water level measurements for each month available for this study. \autoref{fig:map} shows the spatial distribution of altimetry data along the river network. For the short-repeat missions the length of the water level time series is colour-coded.


\newcommand{\Gitter}[4]{
     \draw [help lines,very thin,dashed, color=gray, ystep=100, xstep=1] (#1,#3) grid (#2,#4);
}
\newcommand{\Koordinatenkreuz}[6]{
    \draw[-, >=latex,thick, color=black] (#1,#3) -- (#2,#3) node[right] {#5};
    \draw[-, >=latex,thick, color=black] (#1,#4) -- (#2,#4) node[right] {#5};
    \draw[-, >=latex, thick,color=black] (#1,#3) -- (#1,#4) node[left] {#6};
    \draw[-, >=latex,thick, color=black] (#2,#3) -- (#2,#4) node[left] {#6};
}

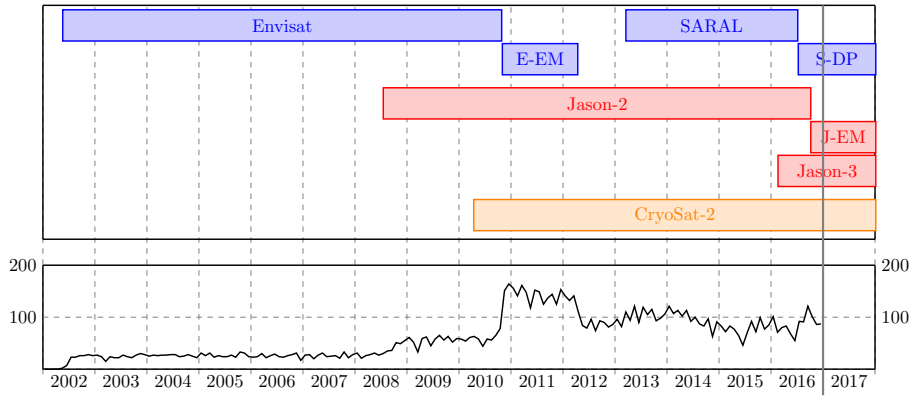
\begin{figure}[ht]
\centering
\begin{tikzpicture}[scale=1, samples=100, >=latex]
    \Gitter{0}{16}{-.1}{7}
    \Koordinatenkreuz{0}{16}{2.5}{7}{}{}
    \Koordinatenkreuz{0}{16}{0}{2.0}{}{}
    \draw[help lines,very thin,dashed, color=gray] (0,1) -- (16,1) node[right] {};
    \draw (0.5,0) node[below]{2002};
    \draw (1.5,0) node[below]{2003};
    \draw (2.5,0) node[below]{2004};
    \draw (3.5,0) node[below]{2005};
    \draw (4.5,0) node[below]{2006};
    \draw (5.5,0) node[below]{2007};
    \draw (6.5,0) node[below]{2008};
    \draw (7.5,0) node[below]{2009};
    \draw (8.5,0) node[below]{2010};
    \draw (9.5,0) node[below]{2011};
    \draw (10.5,0) node[below]{2012};
    \draw (11.5,0) node[below]{2013};
    \draw (12.5,0) node[below]{2014};
    \draw (13.5,0) node[below]{2015};
    \draw (14.5,0) node[below]{2016};
    \draw (15.5,0) node[below]{2017};
    
    \draw (0,1) node[left]{100};
    \draw (0,2) node[left]{200};
    \draw (16,1) node[right]{100};
    \draw (16,2) node[right]{200};

    \node[rectangle, anchor=south west,draw, thick,color=blue,fill=blue!20,minimum width=8.44cm,minimum height=.6cm] (envisat) at (0.367,6.3) {{\centering Envisat}} ;
    \node[rectangle, anchor=south west,draw, thick,color=blue,fill=blue!20, minimum width=1.45cm, minimum height=.6cm] (em) at (8.82,5.65)  { {\centering E-EM}};
     \node[rectangle, anchor=south west,draw, thick,color=blue,fill=blue!20, minimum width=3.305cm, minimum height=.6cm] (saral) at (11.195,6.3)  { {\centering SARAL}};
     \node[rectangle, anchor=south west,draw, thick,color=blue,fill=blue!20, minimum width=1.49cm, minimum height=.6cm] (dp) at (14.51,5.65)  { {\centering S-DP}};
     \node[rectangle, anchor=south west,draw, thick,color=red,fill=red!20,minimum width=8.22cm,minimum height=.6cm] (j2) at (6.53,4.8) {{\centering Jason-2}} ;
     \node[rectangle, anchor=south west,draw, thick,color=red,fill=red!20,minimum width=1.25cm,minimum height=.6cm] (j2em) at (14.75,4.15) {{\centering J-EM}} ;
     \node[rectangle, anchor=south west,draw, thick,color=red,fill=red!20,minimum width=1.88cm,minimum height=.6cm] (j3) at (14.12,3.5) {{\centering Jason-3}} ;
     \node[rectangle, anchor=south west,draw, thick,color=orange,fill=orange!20,minimum width=7.73cm,minimum height=.6cm] (c2) at (8.27,2.65) {{\centering CryoSat-2}} ;
     \draw[-, >=latex,very thick, color=gray] (15,-.5) -- (15,7) node[right] {};
        \makeatletter
    \draw[thick, -]\@@input num_points.txt ;
 \makeatother

\end{tikzpicture}
\caption{Temporal availability of altimetry missions. Data are used in this study until the end of 2016, as indicated by the grey vertical line. The lower panel shows the number of water level observations used in this study for each month.}
\label{fig:data}
\end{figure}

\subsection{In Situ Data}

The results of the multi-mission altimetry approach are validated with  water level data from in situ gauges for the flood season provided by the Mekong River Commission (MRC, \url{http://ffw.mrcmekong.org/}). Gauge data with a daily resolution are extracted from the 1st of June to the 30th of November for the years 2008 to 2016. Locations and names of the gauging stations are shown in \autoref{fig:map}.

\section{Universal Kriging for Multi-Mission Altimetry Combination}
\label{sec:method}
For the incorporation of long or non-repeat orbit missions in a multi-mission approach the universal kriging method (UK) \citep{Delfiner1975,Myers1982} is applied in this study to link all altimetry data together from different tributaries and streams of the river. UK has the advantage that it does not require a constant mean over the data, unlike ordinary kriging (OK) of the  previous multi-mission altimetry study in \citet{Boergens2017}. The constant mean of the data could only be fullfilled for altimetry by reducing the water level observation by their location's mean water level, i.\,e. the topography. While the mean water level of VSs of short-repeat orbit missions (Envisat, SARAL, Jason-2, Jason-2 EM and Jason-3) can be derived directly from the mean of the time series, this is not possible for long or non-repeat missions. To extract the mean water levels from topography models for the reduction is not feasible due to the inaccuracy of these models. Altimetry data from long or non-repeat orbit missions are necessary to acquire additional spatial samples compared with only using short-repeat orbit altimetry. Thus, for the incorporation of altimetric water levels from long or non-repeat orbit missions such as CryoSat-2, SARAL DP and Envisat EM it is a necessity to employ UK. 

Unknown mean water levels or topography along the river are modelled with an unknown linear combination of known functions $\{f_0(\bs x)...f_p(\bs x)\}$ . The water level $Z(\bs s,t)$ observed at location $\bs s$ and at time $t$ can be formulated as: 

\begin{equation}
 Z(\bs s,t) = \sum_{j=0}^{p}f_j(\bs s)\beta_j + \delta(\bs s,t).
\end{equation}

Where $\delta(\bs s,t)$ is the variation in water level at a given location and w.r.t. the mean water level $\sum_{j=0}^{p}f_j{\bs s}\beta_j$ with the unknown parameter $\beta_j$.  We use polynomial B-Splines with a spline degree of three for $f_j(\bs s)$ \citep{Stollnitz1995, Schmidt2015}. 

%
%

The kriging equation for the interpolation of a water level at point $\bs s_0$ and time $t_0$ is 

\begin{align} \label{eq:OKpred}
p(\bs s_0,t_0) = \sum_{i=1}^n \lambda_i Z(\bs s_i,t_i),
\end{align}

with the altimetric water level observations $Z(\bs s_i,t_i)$ at the locations $\bs s_i$ and time $t_i$. In UK, it is of no significance by which altimeter mission the water level $Z(\bs s,t)$ is measured, every water level observation is treated equally in the approach and is only depending on its location $\bs s$ and time $t$. The weights $\lambda_i$ are estimated as
\begin{align}
\bs \lambda=(\lambda_1,\ldots,\lambda_n)=\left( \bs c_U + \mathbf{F}(\mathbf{F}^{\top}\bs{\Sigma_U}^{-1}\mathbf{F})^{-1}(\mathbf{f} - \mathbf{F}^{\top}\bs{\Sigma_U}^{-1}\bs c_U) \right)^{\top}\left(\bs{\Sigma_U}+ \bs{\Sigma}_{alti}\right)^{-1}
\end{align}
where $\bs c_U$  contain the covariances of all observations to the interpolation point and $\Sigma_U$ the covariances among all observations, thus $\bs{c}[i]=C((\bs s_0, t_0),(\bs s_i, t_i))$ and $\bs{\Sigma}[i,j]= C((\bs s_i, t_i),(\bs s_j, t_j))$. $\mathbf{f}$ and $\mathbf{F}$ contain all basis functions evaluated at the interpolation point and all observation locations respectively, thus $\mathbf{f}[j]=\mathrm{f}_{j}(\bs s_0)$ and $\mathbf{F}[j,i]=\mathrm{f}_{j}(\bs s_i)$. $\Sigma_{alti}$ includes the different accuracies of the data (see \mbox{\citet{deMarsily1986}}).

To derive the covariance function between the observations required in UK, we use the non-stationary spatio-temporal covariance model introduced in  \citet{Boergens2017}, which allows the modelling of the inflow of tributaries and of different flow behaviours along the river. This allows to answer the second research question of this study, how data along tributaries can be included. The covariance model consists of independent spatial and temporal components. The temporal component is an exponentially declining covariance model. The spatial component consists of two elements. The first element depends on the distance along the river, if the locations are connected. The second element relates the location of the sub-basins to each other. The river-distance related covariance is a non-stationary covariance model based on the flow between points.  The details of the covariance functions and their parameter estimations can be found in \citet{Boergens2017}.

\section{Results and Validation}
\label{sec:results}

We use UK to interpolate multi-mission time series during flood seasons in the years 2008 to 2016 at the locations of all gauging stations shown in \autoref{fig:map}. It is possible to use any location along the river but the gauging stations are chosen for validation. The temporal resolution of the time series is set to five days based on mean data availability in the study area. The approach allows to interpolate water level time series for the whole year, but in this study only the flood season between June 1st and November 30th is investigated. Due to two reasons, the results of the UK method are presented for the flood season only: First, in situ data is only available for the whole year since 2013 before this only flood season data is provided, therefore validation for the dry season in the years 2008--2012 is not possible. Second, we aim to assess the ability of multi-mission altimetry to quantify extreme water levels, which only occur in the Mekong River during the flood season  \citep{Adamson2009}.

The UK approach is investigated under two situations. In \autoref{sec:timeseries}, we investigate and validate the interpolated water level time series during the flood season. And in \autoref{sec:extreme_events} we demonstrate the ability of the multi-mission altimetry time series to quantify extreme events. 

\subsection{Accuracy of Time Series and Consistency with In situ Data}
\label{sec:timeseries}
To evaluate the influence of tributaries on results, we conduct three different estimations. Water level observations along tributaries may contain valuable information on the main river stream and increase data availability. However, observations of water levels in tributaries are less accurate due to their smaller size.
The three scenarios are: 
\begin{itemize}
 \item S-I: altimetry data for the whole river basin, including tributaries; 
  \item S-II: altimetry data only for the main river stream; 
 \item S-III: altimetry data on the main river stream and major tributaries (see the inlay in \mbox{\autoref{fig:map}}).
\end{itemize}

The third dataset including only major tributaries constitute an intermediate scenario, as not all tributaries are equally important to the main river flow. Major tributaries in S-III are defined by their relative inflow to the total inflow of the river \mbox{\citep{MRC2005}} (see \mbox{\autoref{fig:map}} for the major tributaries). In the second and third scenarios observations of tributaries are assigned a higher covariance factor than observations on the main stream in  $\Sigma_{alti}$ (see \mbox{\autoref{sec:method}}). 

In order to investigate the improvement of the interpolated water level time series due to CryoSat-2 data, we compare the results with a multi-mission approach which only combines the short-repeat orbit data, Envisat, SARAL, and Jason-2, based on ordinary kriging (OK) \citep{Boergens2017}. In this OK multi-mission approach only S-III (data of main river and major tributaries) is adopted. \replaced{The same covariance model as for the UK approach is applied to allow for a better comparison between the approaches. But, the covariance model parameters might change if fitted to the smaller data set of only the short-repeat orbit missions (as done in \citet{Boergens2017}).}{The same covariance model is applied though the model parameter might change with different data basis.}
\added{It should be noted that the OK results presented here are not directly comparable with the results in \citet{Boergens2017} due to two main reasons. First, the time frame is different as only flood season results presented here. During the flood season, the water level exhibits more short-term variations, which deteriorate the accuracy of the resulting time series (see also the discussion on short-termed flash floods in \autoref{sec:discussion}). Additionally, the pre-processing of the altimetry data is different between the two studies. Here an annual signal is not reduced from the data as this is not possible for the long- and non-repeat orbit missions whereas in \citet{Boergens2017} the kriging is done only on the residuals.}

The results are additionally compared to single-mission altimetry results of Envisat/SARAL or Jason-2 VSs. Because of the higher temporal resolution and better data availability of Jason-2, we decide to use a Jason-2 time series instead of a closer Envisat and SARAL time series for the closest VS when the Jason-2 VS is less than 100\,km away. Envisat and SARAL have a data gap in 2011 and 2012, whereas Jason-2 is available for the whole timeframe of the study. VS data are derived from DAHITI (see \mbox{\autoref{sec:data}}). At some stations (Chiang Khan and Chiang Saen), the closest VS only includes Envisat data as SARAL data quality is insufficient. In this case only floods in the years 2008--2010 are monitored by the VS.

In situ data from gauging stations are used to validate the altimetry-based time series. Validation is performed with three performance measures: root mean square error (RMSE), coefficient of determination (squared correlation coefficient; R$^2$) and Nash-Sutcliffe model efficiency coefficient (NSE) \citep{Nash1970}. RMSE measures absolute differences between gauge and altimetry time series, whereas the coefficient of determination is sensible to phase shifts between time series. The maximum value of NSE is one and a value of zero indicates that the time series observes the in situ times series as well as the mean observed water level. NSE values below zero indicate that the mean water level would be a better approximation to the observed in situ time series than the altimetric time series. The validation results are shown in \autoref{tab:results}. \added{The influences of the different missions on the resulting time series depend on the location of the station and are constantly changing for each time step. }

\begin{figure}
 \centering
 \includegraphics[width=\textwidth]{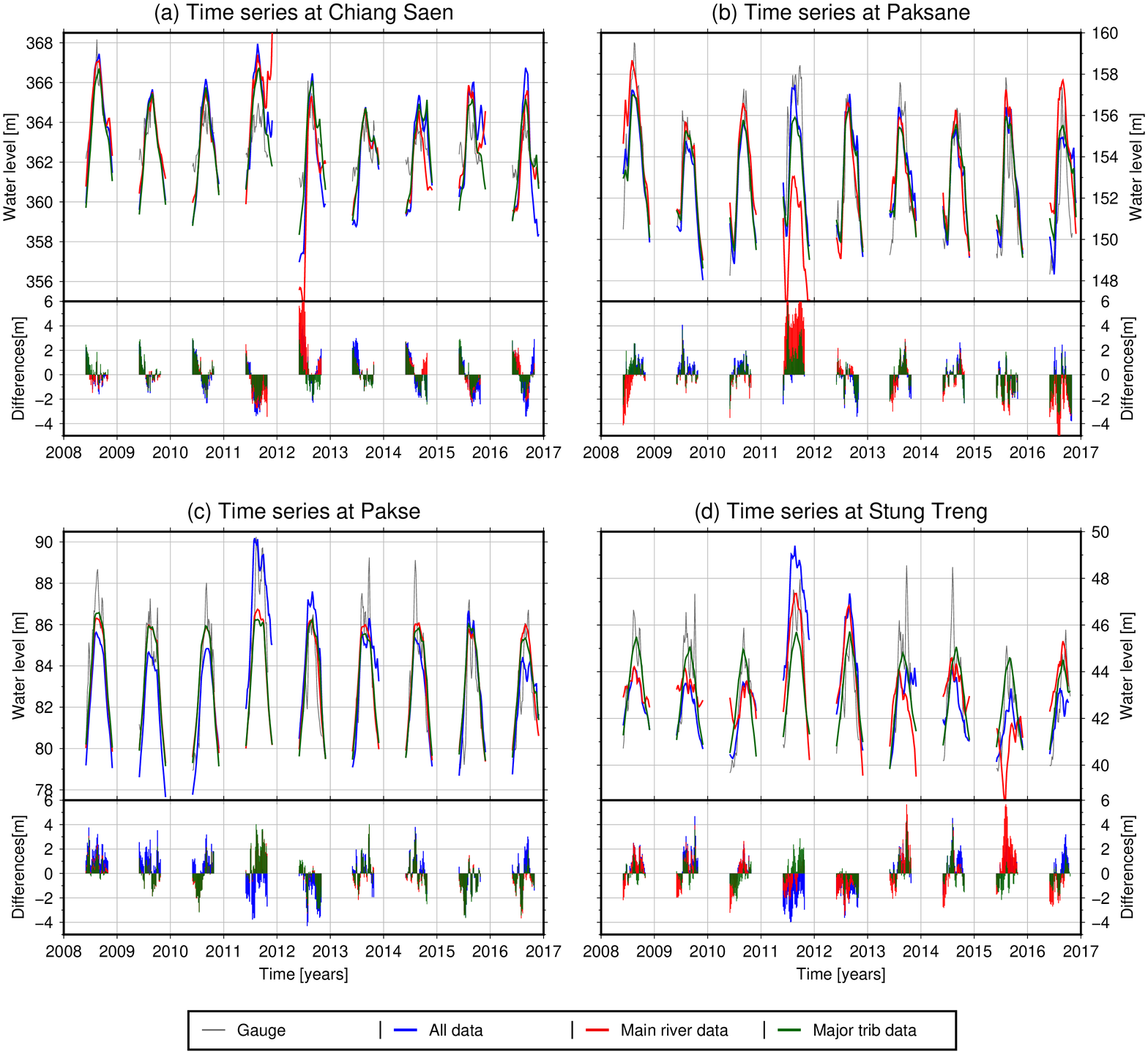}
 \caption{Resulting multi-mission altimetry time series at four stations. Differences between the gauge and altimetric time series (gauge-altimetry) are shown for each station.}
 \label{fig:timeseries}
\end{figure}

At many stations, performances for all three data sets used in UK are similar. S-I show intermediate performance (RMSE range: 1.20\,m--1.67\,m; R$^2$ range: 0.66--0.88; NSE range: -0.97--0.77). S-II perform worst (RMSE range: 1.19--2.22\,m; R$^2$ range: 0.5--0.88; NSE range: -1.03--0.77), while S-III perform best (RMSE range: 1.05--1.75\,m; R$^2$ range: 0.81--0.9; NSE range: 0.02--0.8). The improvements from S-II to S-I or S-III are more innate in terms of R$^2$ than RMSE or NSE. Correlations between the quality of the results and locations along the river are not significant. NSE values do not fall below 0.59 except for Chiang Saen \replaced{Paksane, and Stung Treng stations in S-II (and S-I for Stung Treng).}{and Paksane stations in S-II.} The OK results are inferior to the UK results at all stations except at Chiang Saen, where they are significantly better, and Paksane with similar results. The differences between OK and UK results vary but without any apparent geographical correlation. In general the amplitudes are underestimated in the OK time series. Overall, the UK results are 8\% better in terms of RMSE and 4\% better in terms NSE. The coefficients of determination shows smaller differences.
The results of VSs are inferior to UK results, except at Pakse and Paksane stations, both measured by Jason-2. However, flood behaviour may change over a distance of 100\,km between the gauge and the VS (see differences between Chiang Khan and Vientiane stations in 2008 in \autoref{fig:flood_index_uk}).  Four of the stations (Chiang Saen, Paksane, Pakse and Stung Treng) have a notable different behaviour. The time series of these stations are displayed in \autoref{fig:timeseries}. Differences between the gauge and altimetric time series are given below each station's time series in \autoref{fig:timeseries}. The reasons of these different behaviours will be discussed in \autoref{sec:discussion_results}.

Chiang Saen station (\autoref{fig:timeseries} (a)) is the most northern gauging station included in this study.  The results of S-I and S-II are inferior to the results in S-III at this station. Amplitude is overestimated in all scenarios, leading to low NSE values.   Differences show similar behaviours among years, with decreasing positive differences before the flood peak and negative differences after the flood. The OK time series observes the correct amplitude leading to the better performance.

At Paksane station the time series in S-II shows a distinct offset in 2011  while S-I and S-III remain on the correct level (\autoref{fig:timeseries} (b)). This offset deteriorates the validation measures for S-II significantly.  This may be caused by a single major outlier in the dataset observed by a long or non-repeat orbit mission. Hence, the performance of the OK time series appears to be better as it does not include this error.

In the Pakse station time series (\autoref{fig:timeseries} (c)), only the S-I results accurately reflect the 2011 flood, while the other two scenarios show amplitudes that are too low. S-II and S-III show similar results but fail to quantify inter-annual variations in water levels. However, over the whole time series S-II performs best according to the validation measures in \autoref{tab:results}.

In Stung Treng S-III clearly outperforms the two other scenarios. S-II fails to correctly observe the annual signal before 2011 and in 2015, while S-I underestimates the floods prior 2011 and overestimates the 2011 flood. Only S-II correctly quantifies the 2011 flood. The differences show a recurring pattern each year, but less pronounced than at Chiang Saen.

\begin{table*}[]
\centering
\rowcolors{1}{}{Gray}
\caption{Validation of the multi-mission time series against in situ gauge data. Comparison to multi-mission results with only short-repeat orbit data and single mission altimetry, which is the closest Virtual Station (VS), is given at the end of the table. The parentheses following each station name indicate the distance between the gauge and the VS, and the name of the mission used to measure the VS (E = Envisat, S = SARAL and J2 = Jason-2). RMSE: Root Mean Square Error. R$^2$: coefficient of determination. NSE: Nash-Sutcliffe model efficiency coefficient.}
\label{tab:results}
\resizebox{1.1\textwidth}{!}{
\begin{tabular}{@{}|l|lll|lll|lll|lll|lll|@{}}
\toprule
 \multicolumn{1}{l|}{}& \multicolumn{9}{l|}{\textbf{Multi-Mission of short, long, and non-repeat orbit missions}} & \multicolumn{3}{p{4cm}|}{\textbf{Multi-Mission of short repeat orbit missions}} & \multicolumn{3}{l|}{\textbf{Single-Mission}} \\ \cmidrule(l){2-16} 
 \multicolumn{1}{l|}{} & \multicolumn{3}{l|}{All tributaries (S-I)} & \multicolumn{3}{l|}{Only main river (S-II)} & \multicolumn{3}{l|}{Major tributaries (S-III)} & \multicolumn{3}{l|}{Major tributaries (S-III)} &  &  &  \\ \cmidrule(l){2-16} 
 \multicolumn{1}{l|} {}& RMSE {[}m{]} & R$^2$ & NSE & RMSE {[}m{]} & R$^2$ & NSE & RMSE {[}m{]} & R$^2$ & NSE & RMSE {[}m{]} & R$^2$ & NSE & RMSE {[}m{]} & R$^2$ & NSE \\ \midrule
Chiang Saen (72\,km, E) & 1.66 & 0.76 & -0.97 & 1.69 & 0.73 & -1.03 & 1.24 & 0.81 & 0.02 & 0.58 & 0.89 & 0.78 & 2.64 & 0.45 & -8.86 \\
Luang Prabang (-36\,km, J2) & 1.66 & 0.78 & 0.61 & 1.54 & 0.81 & 0.66 & 1.75 & 0.83 & 0.60 & 1.75 & 0.85 & 0.60 & 1.77 & 0.62 & 0.32 \\
Chiang Khan (12\,km, E) & 1.39 & 0.84 & 0.63 & 1.34 & 0.85 & 0.65 & 1.05 & 0.90 & 0.80 & 1.30 & 0.88 & 0.69 & 3.27 & 0.45 & -1.41 \\
Vientiane (-22\,km, E+S) & 1.2 & 0.87 & 0.76 & 1.19 & 0.87 & 0.76 & 1.31 & 0.89 & 0.74 & 1.39 & 0.90 & 0.70 & 2.43 & 0.51 & -0.43 \\
 Nong Khai  (-56\,km, E+S) & 1.21 & 0.88 & 0.77 & 1.22 & 0.88 & 0.77 & 1.22 & 0.89 & 0.79 & 1.46 & 0.90 & 0.70 & 2.45 & 0.5 & -0.44 \\
Paksane   (41\,km, J2) & 1.31 & 0.86 & 0.73 & 2.22 & 0.65 & 0.3 & 1.44 & 0.86 & 0.72 & 1.31 & 0.91 & 0.77 & 1.28 & 0.88 & 0.67 \\
Nakhon Phanom (-19\,km, E+S) & 1.36 & 0.86 & 0.74 & 1.36 & 0.87 & 0.73 & 1.36 & 0.86 & 0.73 & 1.79 & 0.75 & 0.54 & 2.81 & 0.67 & -1.17 \\
Mukdahan   (42\,km, E+S) & 1.23 & 0.88 & 0.77 & 1.46 & 0.83 & 0.67 & 1.31 & 0.87 & 0.73 & 1.46 & 0.91 & 0.67 & 2.06 & 0.55 & -0.38 \\
Pakse    (-41\,km, J2) & 1.67 & 0.80 & 0.59 & 1.39 & 0.85 & 0.72 & 1.47 & 0.84 & 0.70 & 1.77 & 0.87 & 0.56 & 1.26 & 0.94 & 0.78 \\
Stung Treng (18\,km, E+S) & 1.63 & 0.66 & 0.31 & 1.83 & 0.50 & 0.11 & 1.13 & 0.84 & 0.68 & 1.45 & 0.70 & 0.47 & 1.88 & 0.38 & -0.69 \\ \bottomrule
\end{tabular}}
\end{table*}

\subsection{Monitoring Extreme Events}
\label{sec:extreme_events}
In order to answer the last research question of this study, the inter-annual behaviour of the floods are investigated in this section.
The extent of the flood season is evaluated with a novel flood index, based on the mean differences in each year between the long-term annual signal and observed water level. The flood index $f$ for the year $y$ and the location $\bs s$ is defined as:

\begin{equation}
 f(\bs s, y) = \frac{1}{N}\sum_i Z(\bs s, d_i^y)-\overline{G(\bs s, d_i)}.
\end{equation}

Where $Z(\bs s, d_i^y)$ are observed water levels at the location during the flood season in the year $y$; $d_i^y$ is the $i$th day in the year $y$; and $\overline{G(\bs s, d_i)}$ is the long-term mean gauge reading over all years on a given day $d_i$. \added{A positive flood index indicates an above average water level and vice versa.}

Floods for each year and for each gauging station are evaluated with this flood index for the gauging data itself, the interpolated five-day UK time series, and the nearest VS of a short-repeat altimetry mission. We use the same VSs as in the previous section. We only use results from S-III as these showed the highest performance in \autoref{sec:timeseries}. Gauge data are reduced to the same temporal resolution as the UK time series (five days). The long-term annual mean used for the index is calculated from gauge data and used for the altimetry-based flood indices as well.

\begin{figure}
 \centering
 \includegraphics[width=\textwidth, clip, trim=2cm 2cm 7cm 1cm]{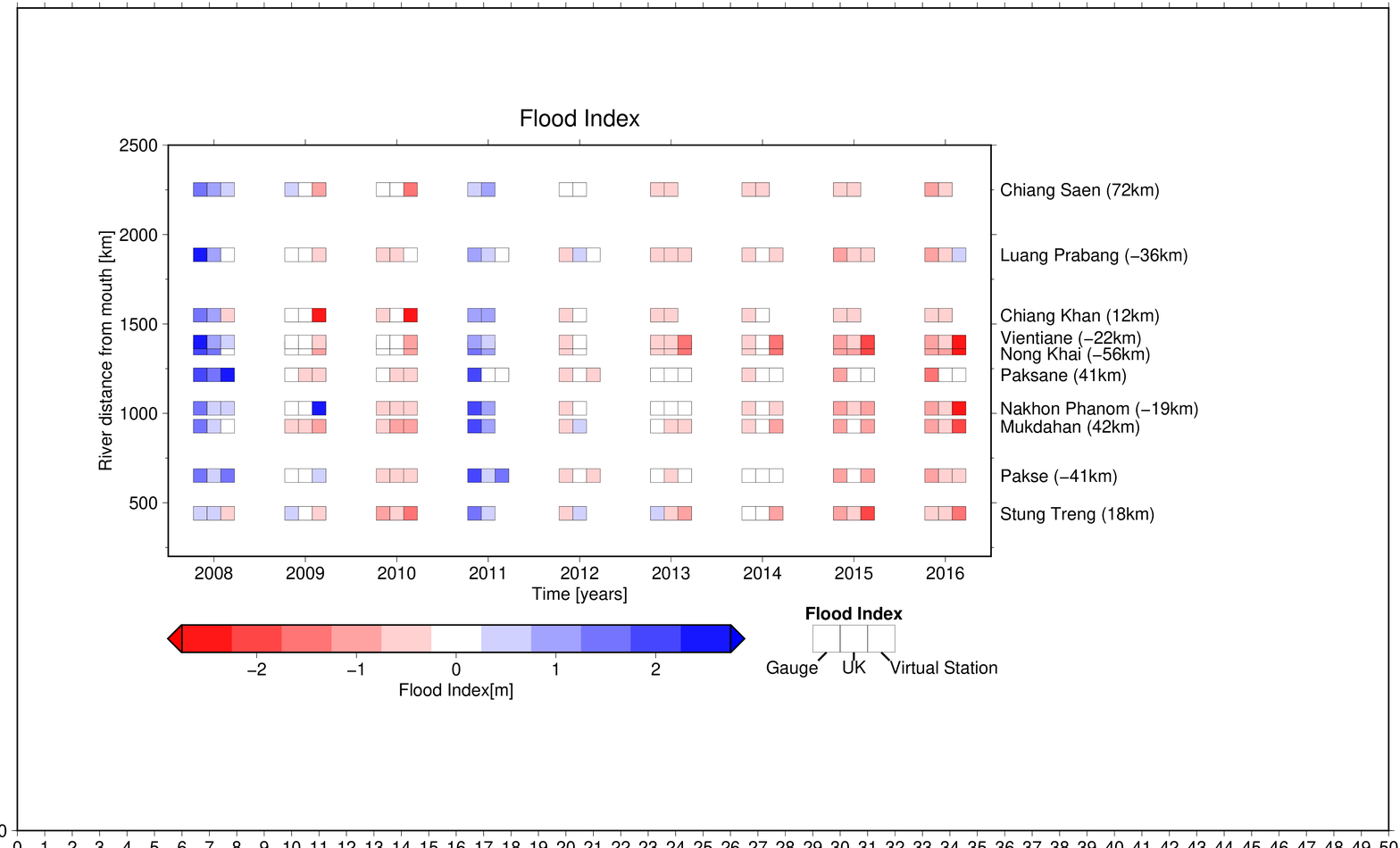}
 \caption{Flood index for all stations from gauge data, UK results, OK results, and VSs. \added{The distance to mouth is computed along a river centreline polygon}}
 \label{fig:flood_index_uk}
\end{figure}

For each year and station, the three flood indices (gauge, UK, and VS) are shown in \autoref{fig:flood_index_uk}. The first square is the flood index for the in situ gauge data and represents the ground truth. The names of the stations are given together with the distances to the closest VS. 

The Mekong Basin was affected by two major flooding events during the study, in 2008 and 2011 (see also \autoref{fig:timeseries}). These extreme events are detected in both the gauge data and UK results. However, the extent of the flooding is underestimated by the UK time series. At Paksane station, flooding in 2011 is  not detected in the UK time series. Only the Jason-2 VSs detect the floodings, whereas many of the Envisat/SARAL VSs fail to detect them. Anomalously low water levels during the flood season, which are called a hydrological drought, in 2015 and 2016 are not detected by all UK time series, but are overestimated by some VSs. Medium flows are observed in 2009, but the VS close to Nakhon Phanom erroneously detected a flooding, and the VS close to Chiang Khan erroneously detected an exceptionally low flow. 

The comparison of gauge and altimetry flood indices can also be evaluated by means of quality measures: The coefficient of determination (R$^2$) between the gauge and UK flood indices is 0.81, and between gauge and VSs is 0.51. The ability to detect floodings or droughts is measured with the Probability of Detection (PoD), while the False Alarm Ratio (FAR) quantifies the amount of false flooding or drought detection. To this end, we define a flooding event if the flood index exceeds 0.5 whereas a drought is observed if the index is below -0.5. These thresholds are chosen such that the known events in 2008, 2011, 2015, and 2016 are correctly depicted as flooding or drought in the in situ gauge data. The PoD is defined as the ratio of correctly observed events by altimetry to the total amount of these events observed by the gauge. On the other hand, the FAR relates the number of false detection of a flooding or drought event by altimetry to the number of all altimetry detected flooding or drought events \citep[chap. 8.2]{Wilks2011}.  The PoD for floodings with UK multi-mission altimetry is 79\% while for droughts it is only 28\%; the FAR is 6\% for floodings but 41\% for droughts.  

For comparison, the OK time series are also capable to observe the inter-annual variations but not as good as the UK results. The coefficient of determination between gauge and OK flood indices is 0.78. OK has a lower PoD for floodings with only 68\% but a better PoD for droughts with 48\%.  On the other hand the FAR is 18\% for floodings and 29\% for droughts. 

Single mission altimetry is less capable of correctly observing floodings (PoD of 38\%) while it performs better for droughts (PoD of 73\%). However, the FAR for droughts is as high as 56\% (28\% for floodings).


\section{Discussion}
\label{sec:discussion}
The following discussion is split into two parts. The first part discusses the findings of the Results section (\autoref{sec:results}), the second part discusses the research questions raised in the introduction.
\subsection{Discussion of Results}
\label{sec:discussion_results}

In \mbox{\autoref{sec:timeseries}} the results at the location of four in situ stations are presented in more detail as they illustrate some specific challenges and problems of our multi-mission altimetry approach.

Two of the four stations, Stung Treng and Pakse, prove in particular the importance of including tributaries in our combination. In Stung Treng the largest of the left bank tributary system of Se Kong, Se San, and Sre Pok joins the Mekong. It accounts for more than 25\% of the total runoff of the river basin.  Ignoring this tributary system negatively affects the time series (\mbox{\autoref{fig:timeseries}} (d)). However, including minor tributaries with the same weight as the Se Kong, Se San and Sre Pok system, degraded the resulting time series. The right bank tributary Nam Mun flows into the main stream a few kilometres upstream from Pakse. This tributary has a high yearly inflow in relation to the main river flow. However, we are unable to incorporate data from this river because the Pak Mun Dam blocks the Nam Mun 5\,km upstream from the confluence. All data above this dam are unusable in UK as the covariance models employed cannot include the flow behaviour across a dam. Therefore the inflow of the Nam Mun cannot be incorporated into the estimation of the time series. Prior to 2011, few datapoints are available around Pakse as only one Jason-2 VS is situated close-by. With few datapoints, the UK approach can only interpolate the mean annual signal. All this together hinders the ability of the multi-mission time series in Pakse to quantify the inter-annual variations of the flood season. This is also visible in \mbox{\autoref{fig:flood_index_uk}}. Over all stations, the inclusion of only the major tributaries improved the results.

Chiang Saen station is the station most influenced by the surrounding topography. There fewer datapoints are available in the vicinity and in lower data quality due to small rivers ($<$200\,m in width). The overestimation of the amplitude is probably caused by the mountainous area where the river mostly flows through narrow gorges but the gauging station is located at a wider part of the valley. This illustrates the problem of combining water levels rather than discharge along the river. The former is directly influenced by rapidly changing topography and thus deteriorates the combination. On the other side, auxiliary data is required to derive discharge values from altimetry. With the multi-mission approach we assume a strong correlation between the water level and discharge which clearly depends on the topography.  The topography problem is most pronounced in the river reach upstream of Vientiane, and in the upstream regions of the left-bank Laotian tributaries.  Chiang Saen is also the only station where the results of OK are significantly better than UK. As can be seen in \mbox{\autoref{fig:map}} three VSs are in the vicinity of the stations and thus are governing the result. Though, the closest VS has a \replaced{too}{to} high amplitude due to the topography and an overall poor performance (see \mbox{\autoref{tab:results}}) the two other VSs, located along a tributary, observe the correct amplitude. \replaced{Especially  Jason-2, available for the whole time span of the study, has a good data quality which leads to the high quality of the OK time series.}{Especially, the Jason-2 time series spanning the whole time of the study has a high data quality which leads to an higher in the quality of the OK time series.}

\begin{figure}
 \centering
 \includegraphics[width=\textwidth]{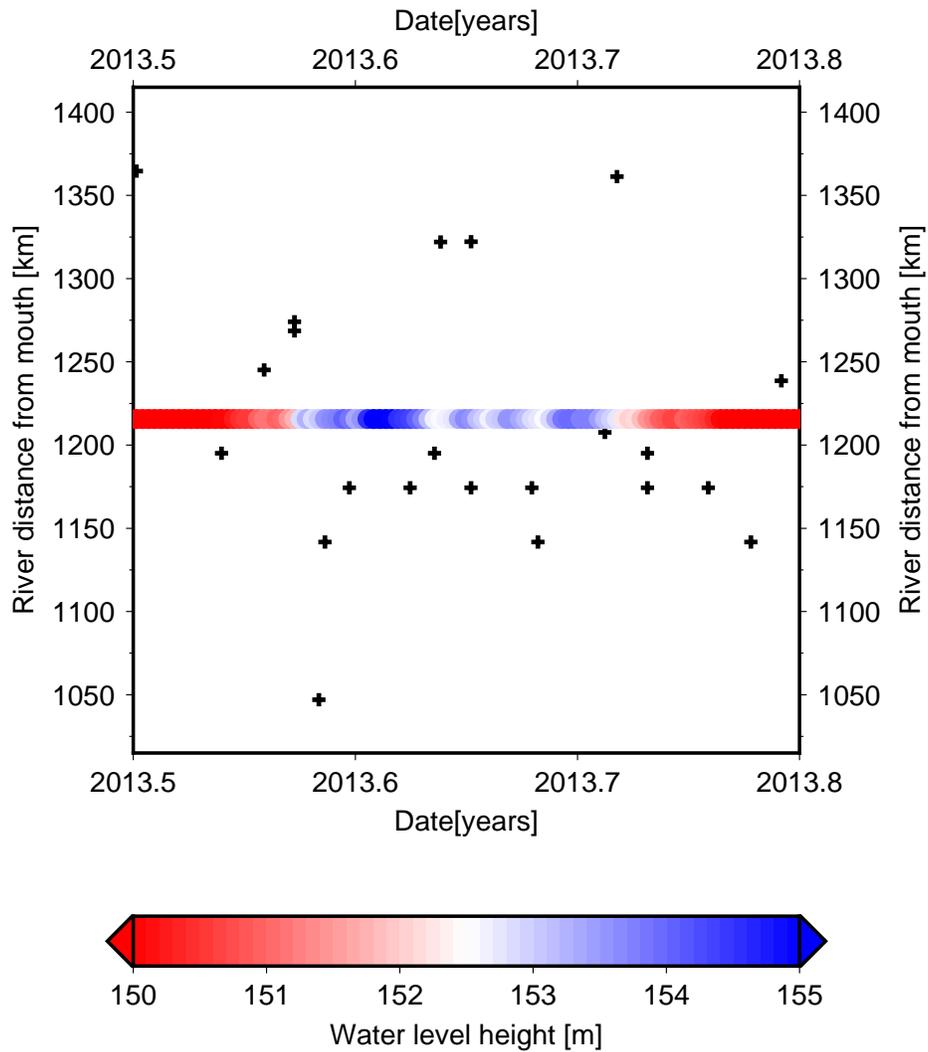}
 \caption{Water levels at Paksane gauging station in 2013. Surrounding altimetric water level observations are shown with black crosses. Flood peaks were not detected by any of the altimetric observations and thus could not be observed in the multi-mission time series. }
 \label{fig:gauge_extreme}
\end{figure}

The multi-mission approach is not always able to identify the main peak of the flood or flash floods. The main peak of the flood lasts for a few days to a month; flash floods are even shorter. The flood peak can only be observed in the multi-mission time series if a water level measurement is available in the vicinity. In \mbox{\autoref{fig:gauge_extreme}} in situ data at the gauging station Paksane are shown (heights colour coded). The available altimeter observations in the surrounding 200\,km of Paksane station are shown as black crosses (\mbox{\autoref{fig:timeseries}}). No altimeter observations are available during the peak of the main flood ($\sim$ 2013.60--2013.62 or 7th--14th of August), therefore the flood peak could not be observed in the multi-mission time series (\mbox{\autoref{fig:timeseries}}). The flash floods following the main flood occur rapidly and could not be detected in the time series (\mbox{\autoref{fig:timeseries}} (b)).

The more data is available, the less critical are outliers as they are evened out. For example, in Paksane the S-II results observe a clearly wrong signal in 2011 while the results with tributaries included (S-I and S-III) are significantly better.

The two stations Chiang Saen and Stung Treng show distinct patterns in the differences between time series and gauge which indicate a phase shift between the two signals. In Chiang Saen the multi-mission time series observes the flood later than the gauge while in Stung Treng the flood is observed earlier. Both stations
are located at the northern and southern border of the study area which leads to an uneven data distribution in the vicinity of the station. In Chiang Saen more downstream data probably lead to the phase shift in the water levels towards a later flood peak. In Strung Treng the main flood peak is observed too early. However, the effect is less pronounced in Stung Treng than at Chiang Saen station.  Stung Treng station is located further from the border of the study area and the data are more evenly distributed around the station.

The inter-annual variability is correctly observed with respect to its trend, but its amplitude is underestimated. Thus, an increase of the threshold used to define a flooding or drought would lead to a lower PoD of both events. The underestimation of the amplitude also leads to the very low FAR which would further decrease with a higher threshold. Over 80\% of the floodings are correctly observed with very few false alarms but the majority of the droughts are not correctly observed and quantified. The underestimation of the floodings can be explained with the not-observable main flood and flash floods (see discussion above), however this does not explain the unquantified droughts in 2015 and 2016.

\subsection{Discussion of Research Questions}
In the Introduction three major research question of this study were raised. In the following we discuss the answers: 

\begin{itemize}
 \item \textbf{What is the benefit of altimetry missions on long and non-repeat orbits for river flood monitoring? }
\end{itemize}
The long and non-repeat orbit missions, most importantly CryoSat-2, provide a dense spatial resolution of river observation and help to close the data gap between Envisat and SARAL. The benefit of these missions is shown in the comparison between the UK results and the OK results, when the latter only used data of short-repeat orbit missions. The interpolated time series with all data have a higher agreement with the in situ gauge time series. More importantly, the probability of a correct flooding detection is 10\% higher with the results including the long and non-repeat orbit missions. \replaced{As the combination based only on short-repeat orbit missions tends to underestimate the amplitude of the seasonal signal, the drought detection performs better with this combination.}{The better drought detection with a combination of only short-repeat orbit missions can be explained by an overall to low estimated amplitude of this combination.}  Thus, including long or non-repeat orbit missions improves the observation of inter-annual flood changes and the detection of floodings but not for droughts.

However, the water level observations of the long or non-repeat orbit mission are more difficult to check for outliers beforehand than the time series of short-repeat orbit missions as the time series cannot be checked for a stable annual signal. Thus uncorrected erroneous data points can distort the results of the multi-mission time series (see discussion above regarding Paksane).
 
\begin{itemize}
  \item \textbf{How can measurements along tributaries be included? How do they influence the quality of water level interpolation? }
\end{itemize}
The multi-mission approach of this study is able to incorporate the data along tributaries to estimate water level time series along the main river. It is shown in the results, that including the major tributaries improves the interpolation for most locations and is in fact necessary for locations close to the confluence with large tributaries (see above the discussion regarding Pakse and Stung Treng). Of ten stations, Chiang Saen, Paksane, Pakse, and Stung Treng are closest located to a confluence and demonstrated the importance of incorporating of tributary data. Luang Prabang is located close to a confluence as well, but here the quality of the time series did not improve in terms of RMSE and NSE (it does in terms of R$^2$). It is possible that some of the data of the major tributary joining the Mekong River upstream of Luang Prabang is faulty. All other stations are less influenced by the tributary data as they are further away from confluences. The influence of tributary data on the interpolated time series does not depend on the location of the station along the river. However, not  all tributaries provide valuable information for the multi-mission approach which could be seen in the results with all tributaries included. 

Without tributaries, the inter-annual flood variations are by far less well observed than with tributaries included. The coefficient of determination is reduced to 0.56 between gauge and multi-mission time series without tributaries while the PoD reduces and the FAR increases. This shows the importance of the information gathered along tributaries to correctly quantify the floods of the main river. 

In the multi-mission approach not water volume but water level is combined. This is the main problem  for the inclusion of tributaries. The water level of a tributary does not influence the water level in the main river with a one-to-one relationship. In the multi-mission approach, this is partly accounted for by down-weighting data from the tributaries. The confluence of a major tributary changes the water level of the main river, while the influence of smaller tributaries is limited. But they might introduce further uncertainty in the estimates due to inferior data quality.

\begin{itemize}
\item \textbf{How well can multi-mission altimetry quantify inter-annual flood variations? Which are the most important factors influencing their accuracy and reliability? }
\end{itemize}
The ability of multi-mission altimetry to observe the inter-annual flood variations is clearly better than in the case  of single-mission altimetry. Overall, UK multi-mission altimetry is underestimating both the floodings and drought amplitude with better results for the floodings. The temporal resolution of multi-mission altimetry is higher than for single-mission altimetry, but it is still not possible to observe every flood peak in the basin. Thus, for some years the full extent of the flood cannot be quantified correctly. In some years the flood maximum is only observed by one observation which makes a quantification of the flood extent vulnerable to erroneous outliers in the data set. 

The distribution of the data around the station has significant influence on the ability to observe inter-annual variations. The three stations Chiang Khan, Vientianne and Nong Khai show the best results for the flood index with a coefficient of correlation to the gauge flood index above 0.9. This is probably caused by the higher data availability and quality in this river reach. Here, the river flows in West-East direction which leads to an almost orthogonal intersection of river and satellite tracks which improves the water level observation. The lowest agreement between multi-mission and gauge flood variation observation is found at the two stations Chiang Saen and Stung Treng which are each located at the edge of the study area with less available data.

\section{Conclusion}
\label{sec:conclusion}

This study demonstrated the benefit of multi-mission altimetry including CryoSat-2 and other long or non-repeat mission data for the observation of flood events along the Mekong River. Altimetric water levels were dispersed over the whole river basin, which allowed for more continuous monitoring of the river independently from in situ gauging stations. With the proposed UK approach,  altimetric data from all available missions could be combined along the course of the main river and tributaries. 

By applying the UK approach this study combined  altimetry observations of short-repeat orbit missions with data from long or non-repeat orbit missions (CryoSat-2, Envisat EM, and SARAL DP). Especially, CryoSat-2 (SAR and LRM) have made a valuable contribution since 2010 by providing a denser spatial and temporal coverage of data along the Mekong River. This also helped to close the data gap between the end of the Envisat mission in 2010 and the launch of SARAL in 2013. With the available data, we could reach a spatial resolution of a few kilometres at a temporal resolution of up to five days. With the additional data we were able to observe the inter-annual flood variation better than with a combination of only short-repeat orbit missions. 

In this study not only data from the main stretch of the river were combined but also tributaries were included. Data from the tributaries, especially the major ones, increased data availability and incorporated valuable information for determining water levels along the main river stream. Three scenarios were tested in the estimation of the multi-mission time series (all tributaries; only major tributaries; no tributaries). The scenario including only major tributaries, delivered the best results. The RMSE ranges between 1.05--1.75\,m, and the coefficients of determination between 0.81--0.9. The inflow of major tributaries altered the flow of the main river. Thus, without the inclusion of this information, the water level interpolation of the main river deteriorated after the confluence.

This study investigated the inter-annual behaviour of the resulting water level time series. Multi-mission altimetry allowed for the observation of changes during the flood season on  basin scale over many years. The floodings of 2008 and 2011 in the Mekong River Basin, as well as the two anomalous dry flood seasons in 2015 and 2016, were accurately reflected by the multi-mission altimetry time series. In contrast, single mission altimetry may be able to observe floodings and droughts only in a limited number of cases.

Although the temporal and spatial resolution of altimetry data were comparatively high, they were insufficient to detect every peak of the main flood. Flash floods often remained undetected, as the flood peak was too short to be measured in the vicinity of the station. This hindered the correct quantification of the flood by means of multi-mission altimetry. However, in ungauged river basins where no reliable in situ data are available flood observations based only on satellite data can be very valuable.

With the inclusion of future and additional missions (Sentinel-3A, launched 2016, Sentinel-3B, launched 2018, Jason CS/Sentinel-6, planned for 2020 and SWOT, planned for 2021) data availability will further increase, and along with that the ability to detect, predict, and forecast floods and flash floods.

Moreover, the flexibility of UK also allows for the incorporation of other data sets to estimate water levels, such as in situ gauge data and precipitation data. Including such datasets should improve the abilities of UK to quantify floods and flash floods and especially improve the monitoring abilities of the approach.

\vspace{6pt}

\subsection*{Acknowledgements}
Johannes Lucke helped with the processing of the non-repeat orbit missions. The data of the short-repeat orbit missions were taken from DAHITI with Christian Schwatke doing most of the processing. Michael Schmidt helped developing the B-Splines used in the UK approach.

The altimeter observations and geophysical corrections are taken from Open\-ADB (http://openadb.dgfi.tum.de). The altimeter missions are operated and maintained by  ESA  (Envisat, CryoSat-2),  ISRO/CNES  (SARAL/AltiKa),  and  CNES (Jason-2, Jason-3). 

This work was supported by the German Research Foundation (DFG) through the TUM International Graduate School of Science and Engineering (IGSSE).

\bibliographystyle{model2-names.bst}\biboptions{authoryear}
\bibliography{literature}

\end{document}